\documentclass[]{fairmeta}
\usepackage{booktabs}
\usepackage{multirow}
\usepackage{graphicx}
\usepackage{hyperref}
\usepackage{xspace}
\usepackage{amsfonts}
\usepackage{wrapfig}
\usepackage{xcolor}
\usepackage{dsfont}
\usepackage{amsmath,amssymb}
\usepackage[ruled,vlined]{algorithm2e}
\DontPrintSemicolon
\SetAlgoLined
\LinesNotNumbered
\SetKwInput{KwIn}{Input}
\SetKwInput{KwOut}{Output}
\SetKwInput{KwParams}{Params}
\SetKwProg{Fn}{Function}{}{}
\usepackage{graphicx}
\usepackage{booktabs}
\usepackage[dvipsnames]{xcolor}
\usepackage{dsfont}
\usepackage{wrapfig}
\usepackage{multicol}
\usepackage{cleveref}
\definecolor{myblue}{HTML}{61CBF4}
\definecolor{mypink}{HTML}{E59EDD}
\definecolor{myred}{HTML}{D14A4A}
\definecolor{darkgreen}{RGB}{0,100,0}
\definecolor{burgundy}{RGB}{128,0,32}

\title {Spectral Subsurface Scattering from RGB via Biophysical Skin Inversion}

\author{Carlos Aliaga}
\author{Adrian Jarabo}

\affiliation{Meta Reality Labs Research}

\abstract{
In this paper we present a spectral optical inversion for skin for path tracing-based rendering of subsurface scattering. 
Skin is a complex multilayered medium, with appearance determined by the mixture of biophysical chromophores. However, current methods rely on medium homogeneization, with optical parameters obtained via albedo inversion from a reflectance texture and hand-tuned scattering distance and anisotropy. This results into significant art-skilled manual labor for authoring, and an inaccurate scattering profile for skin. 
To solve these problems, we generalize existing albedo inversion techniques, and propose a framework that predicts full-spectral skin scattering parameters from a single RGB diffuse albedo. Our method builds upon a new mixture-of-media representation, that approximates the aggregated multilayered appearance of skin by mixing the aggregated scattering of three uncorrelated media. We train a chained neural decoder that maps RGB diffuse albedo to the optical properties of the mixture of media, including anisotropy, scattering radius and scattering albedo. Then, we show this mixture can be used in a random-walk-based path tracer with minimal modifications, by simply randomly selecting the medium to traverse. 
}\label{abstract}

\date{\today}
\correspondence{Carlos Aliaga \email{carlos@aliagabadal.com}}

\begin{document}

\maketitle

\begin{figure*}[t!]
\centering
  \includegraphics[width=\textwidth]{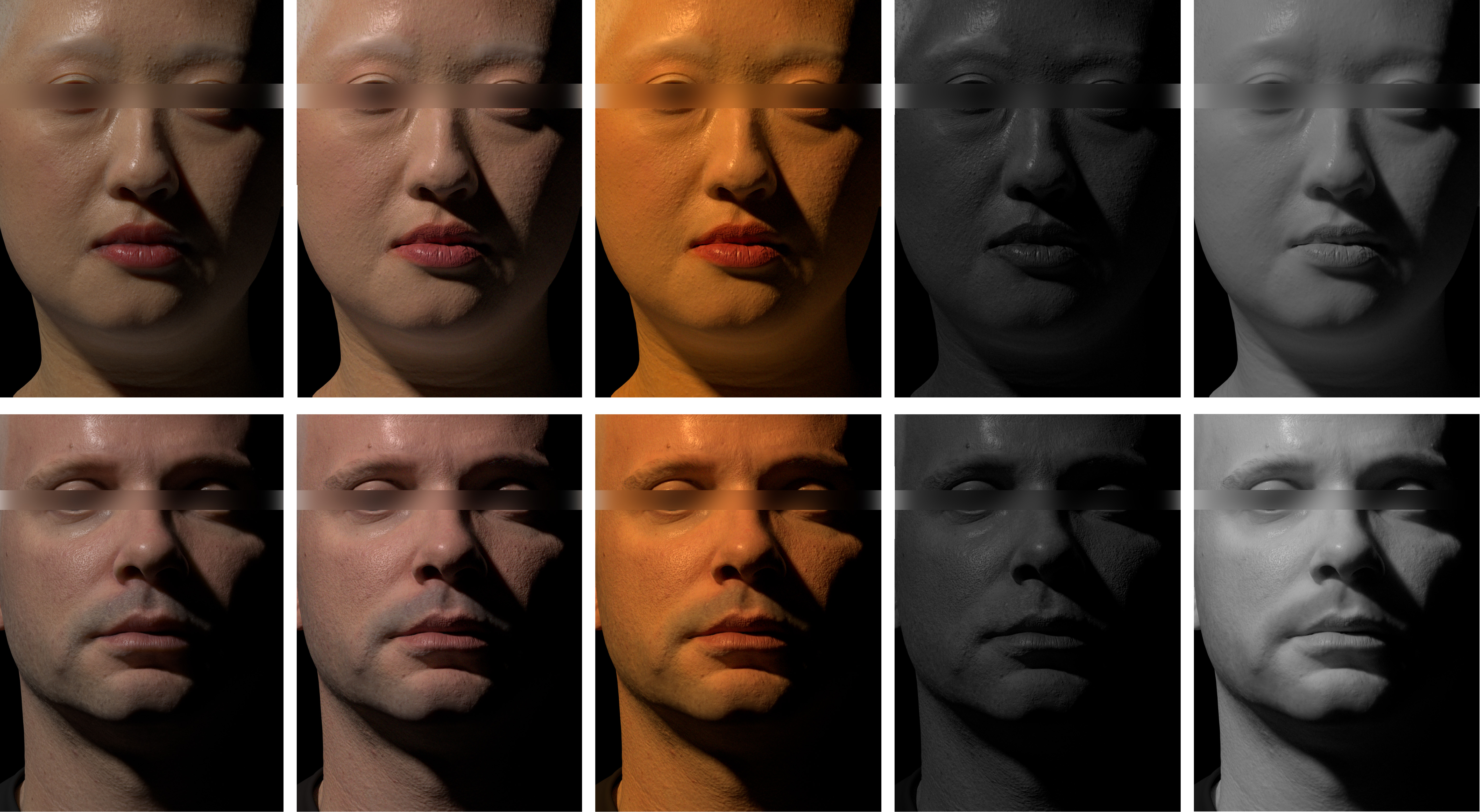}
\caption{Given an RGB skin albedo, our method predicts per-wavelength
subsurface scattering parameters that produce physically correct spectral
translucency for any skin tone.  No prior method predicts per-wavelength
translucency; previous inversion methods \citep{chiang2016practical,wrenninge2017path} use a single artist-set scattering distance
for all wavelengths and skin tones.  Two subjects rendered with side lighting:
(a)~RGB baseline with Wrenninge´s albedo inversion and manually adjusted parameters,
(b)~our method in the visible band (Illuminant~E),
(c)~our method in the visible band (Illuminant~A),
(d)~our method in the UV (250--380\,nm, flat spectrum), and
(e)~our method in the NIR (780--1000\,nm, Illuminant~A).
The dramatic spectral variation in translucency---near-opaque in UV,
highly translucent in NIR---is captured automatically from a single
RGB albedo. Note that while the biophysical model is trained on skin tissue, the parameter ranges extend well beyond typical values for normal skin, allowing the model to accommodate non-skin regions, consistently estimating minimal epidermal thickness for areas such as lips, though the five-parameter model alone cannot fully capture their distinct structure, composition and transport.}
  \label{fig:teaser}
\end{figure*}

\section{Introduction}\label{sec:introduction}

Human skin is a complex scattering multilayered structure, which appearance dominated by internal subsurface scattering within its layers. This scattering is responsible of skin color, but also of its translucent diffuse appearance, specially visible under sharp or back-lighting. 
Subsurface scattering in skin is dominated by its internal structure and by the presence of chromophores with concentrations varying at different depths: Melanin increases epidermal absorption (darkening colour) and reduces the photon mean free path (decreasing translucency); while blood concentration modulates dermal absorption and wavelength-dependent transport distances.

Thus, photorealistic rendering of human skin relies on accurate simulation of subsurface light transport. In modern production path tracers this is done by path tracing an homogeneized idealized scattering medium, parametrized parametrized by a single-medium optical parameters (single scattering albedo $\alpha$ , scattering distance $d$ or equivalently the extinction coefficient $\sigma_t$, and phase function anisotropy $g$). In practice, these parameters are artist-driven, with anisotropy and scattering distance preseted or authored manually by an artist, and single-scattering given by means of \emph{albedo inversion} \citep{chiang2016practical,wrenninge2017path} from a given skin color. 

While this is convenient for authoring general media, this introduces two fundamental issues in the context os skin: First, it decorrelates scattering distance and anisotropy from the skin color. While this works for general media, the manifold of skin appearances is limited by its existing combinations of chromophores, which result on a limited color-dependent range of scattering distances and anisotropies. Moreover, this previous approach assumes that a single homogeneized medium is sufficient to represent the complexity of the aggregated layers.  

In this work we depart from these two assumptions and present a new inversion procedure for obtaining all the \textit{spectral} optical parameters of skin from a RGB skin color image, so it can be used in a modern production renderer. Based on an accurate biophysical skin model \citep{aliaga2023hyperspectral}, we run a large battery of 25,000 skin tones simulations, obtaining both the spectral and RGB skin color and the spectral spatially-resolved diffusion profile. We observe that a fitted single medium is not expressive-enough to accurately model both reflectance and subsurface profile: We instead propose to model skin as a weighted-mixture of $K$ independent media. 

Then, inspired in the previous work of \citet{aliaga2023hyperspectral} we train end-to-end a neural decoder that, from an RGB skin reflectance color,  outputs the full set of spectral optical parameters and mixture weights for a $K$-components mixture of media. This neural decoder is a fast (<1 ms per 4K map) drop-in replacement for artist-set scattering, which generalizes to unseen skin tones and real captured face textures. Finally, we integrate our mixture of $K$-media representation in a path tracer with random walk-based subsurface scattering. 

To the best of our knowledge, this is the first method that extracts the whole set of spectral ready-to-render optical parameters for skin from a single RGB image, which focuses on representing skin beyond merely its reflectance color.


\section{Related Work}\label{sec:related}

\paragraph{Subsurface scattering models for skin.}
While there were some early seminar works using diffusion for rendering skin \citep{stam2001illumination}, the dipole-based seminar work on  BSSRDFs by \citet{jensen2001practical} was the first practical model for translucent materials. Follow up work generalized it to finite multiple layers such skin \citep{donner2005light}, which was later approximated by a sum of Gaussians by \citet{deon2007efficient} in the context of real-time rendering. Our $K$-components mixture draws inspiration of d'Eon's representation, though in the context of path-traced subsurface scattering. Follow up work accelerated d'Eon's model \citep{JimenezSIGA2010,jimenez2015separable} becoming a standard in real-time graphics, while a principled sum-of-Gaussians representation was proposed by \citet{deon2007efficient} in the context of off-line rendering. Later, \citet{christensen2015approximate} proposed an empirical profile model, which has been widely-adopted in production. 
More recently, most off-line production renderers switched to random-walk-based subsurface scattering \cite{christensen2018renderman,fascione2018path}, following the approach proposed by \citet{chiang2016practical}, and incorporating variance-reduction techniques including Dwivedi sampling \cite{kvrivanek2014zero,meng2016improving} and hero-wavelength tracking \citep{wilkie2014hero}. Both diffusion and random-walk based methods require per-channel single scattering albedo and scattering distance parameters. 

\paragraph{Albedo inversion.}
In order to drive the optical parameters required for subsurface scattering, the common approach is to use a texture encoding the surface reflectance (its color) and to obtain optical parameters from it. \citet{chiang2016practical} mapped the surface color to the single scattering albedo, assuming fixed anisotropic scattering and given scattering distance, while \citet{wrenninge2017path} extended it to anisotropic media. In both cases, the scattering distance and anisotropy are free parameters, which requires manual setting.

\paragraph{Biophysical skin models.}
Several models of varying complexity have been proposed to model skin in a biophysical manner: \citet{tsumura1999independent,tsumura2003image} proposed a simple model accounting for spatial distributions of melanin and hemoglobin. Later, \citet{donner2006spectral} introduced an advanced two-layers model, which was later extended to spatially-varying \citep{donner2008layered,chen2015hyperspectral} and time-varying \citep{JimenezSIGA2010,IglesiasEG15} concentration of biophysical parameters. Our work builds upon these works to construct the baseline skin model. More recently, several works tackled the problem of extracting the biophysical parameters from a single skin color image \citep{gitlina2020practical,aliaga2023hyperspectral}. Our work uses a similar approach, but moves it closer to a production setting, by returning ready-to-use optical parameters for rendering in a path tracing setting.

\section{Biophysical Skin Model}\label{sec:model}


We adopt a two-layer biophysical skin model similar to \cite{aliaga2023hyperspectral}. We use this model to simulate a database of spectral scattering profiles and reflectance; this serves as the source of truth relating biophysical properties with emerging scattering behavior. 

\begin{figure}[t]
\centering
\includegraphics[width=0.7\columnwidth]{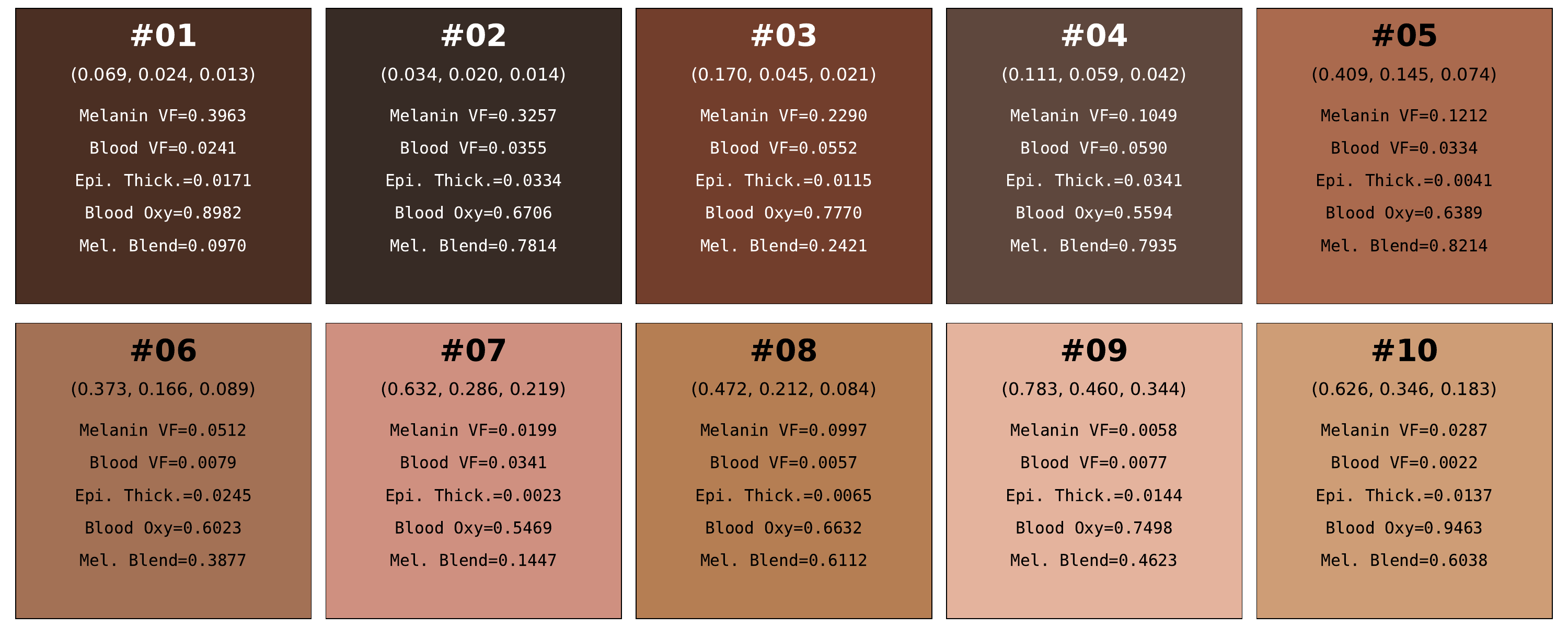}
\caption{Simulated colour swatches for the 10 representative skin tones used
throughout this paper, spanning the full range of human skin diversity from
very light ($f_\text{mel} = 0.001$) to very dark ($f_\text{mel} = 0.65$).
}
\label{fig:swatches}
\end{figure}

\begin{figure}[t]
\centering
\includegraphics[width=\textwidth]{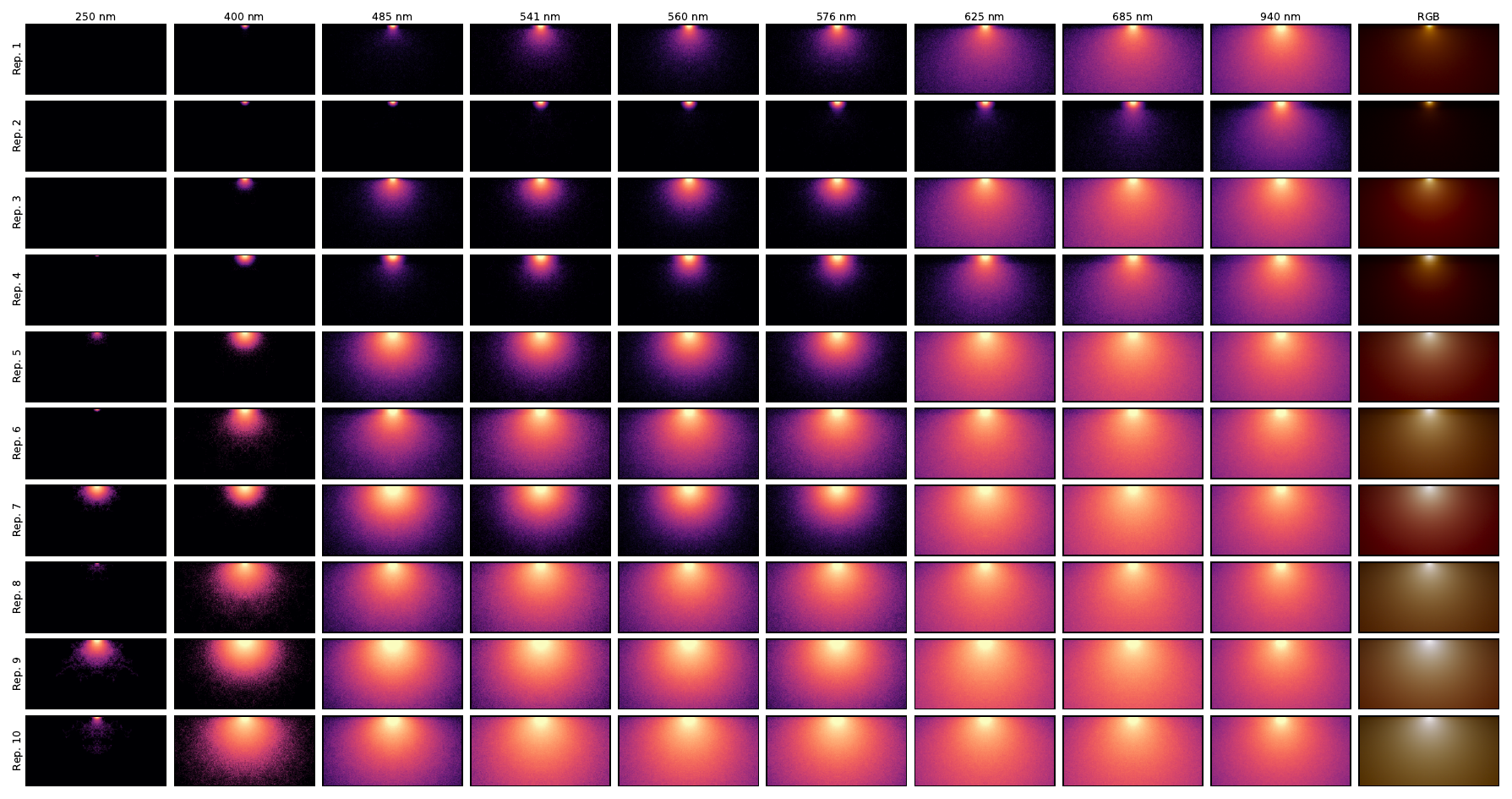}
\caption{Per-wavelength subsurface scattering cross-sections for 10
representative skin tones (light to dark, top to bottom).  Each column shows
a wavelength from UV (250\,nm, opaque) through visible to NIR (1000\,nm,
highly translucent).  The spectral and skin-tone dependence of translucency
motivates our per-wavelength prediction approach (Fig.~\ref{fig:cross_sections}).
}
\label{fig:cross_sections}
\end{figure}

\paragraph{Tissue structure.}
Skin is modelled as a two-layer scattering medium:
An epidermis ($50$--$350\;\mu$m thick, melanin-dominated absorption,
Jacques scattering with $g(\lambda) = 0.62 + 0.00029\,\lambda$~nm) on top of a semi-infinite dermis (with main chromophores haemoglobin, bilirubin, $\beta$-carotene, water). 
We fix basal concentrations for bilirubin, $\beta$-carotene and water, and each skin tone is thus defined by five biophysical properties:
\begin{equation}\label{eq:params}
  \mathbf{p} = (f_\text{mel},\; f_\text{blood},\; t_\text{epi},\;
                \text{S}_{\text{O}_2},\; b_\text{mel}),
\end{equation}
with $f_\text{mel}$ and $f_\text{blood}$ the melanin and hemoglobin concentration, $t_\text{epi}$ the thickness of the epidermis, $\text{S}_{\text{O}_2}$ the blood oxigenation level, and $b_\text{mel}$ the ratio between eumelanin and pheomelanin. We select 10 representative skin tones spanning the full range of human diversity for all figures throughout this paper (Fig.~\ref{fig:swatches}). The spectral subsurface scattering cross sections for such tones are shown in Fig. \ref{fig:cross_sections}. 

\begin{figure}[t]
\centering
\newlength{\figLeftW}\setlength{\figLeftW}{0.38\columnwidth}
\newlength{\figRightW}\setlength{\figRightW}{0.62\columnwidth}

\begin{minipage}[t]{\figLeftW}
    \centering
    \includegraphics[width=\textwidth]{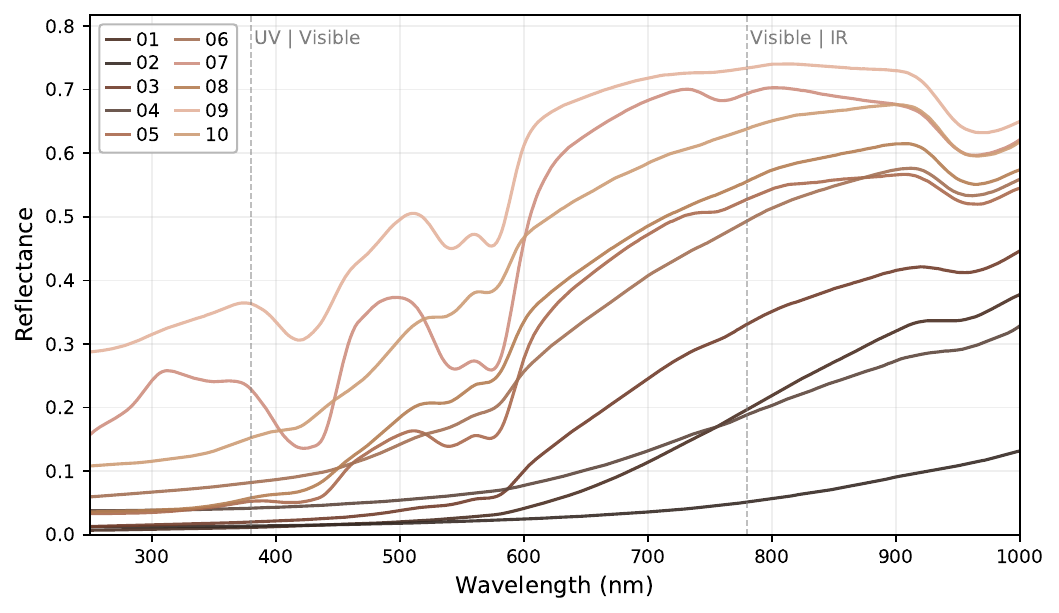}
\end{minipage}%
\hfill
\begin{minipage}[t]{\figRightW}
    \centering
    \includegraphics[width=\textwidth]{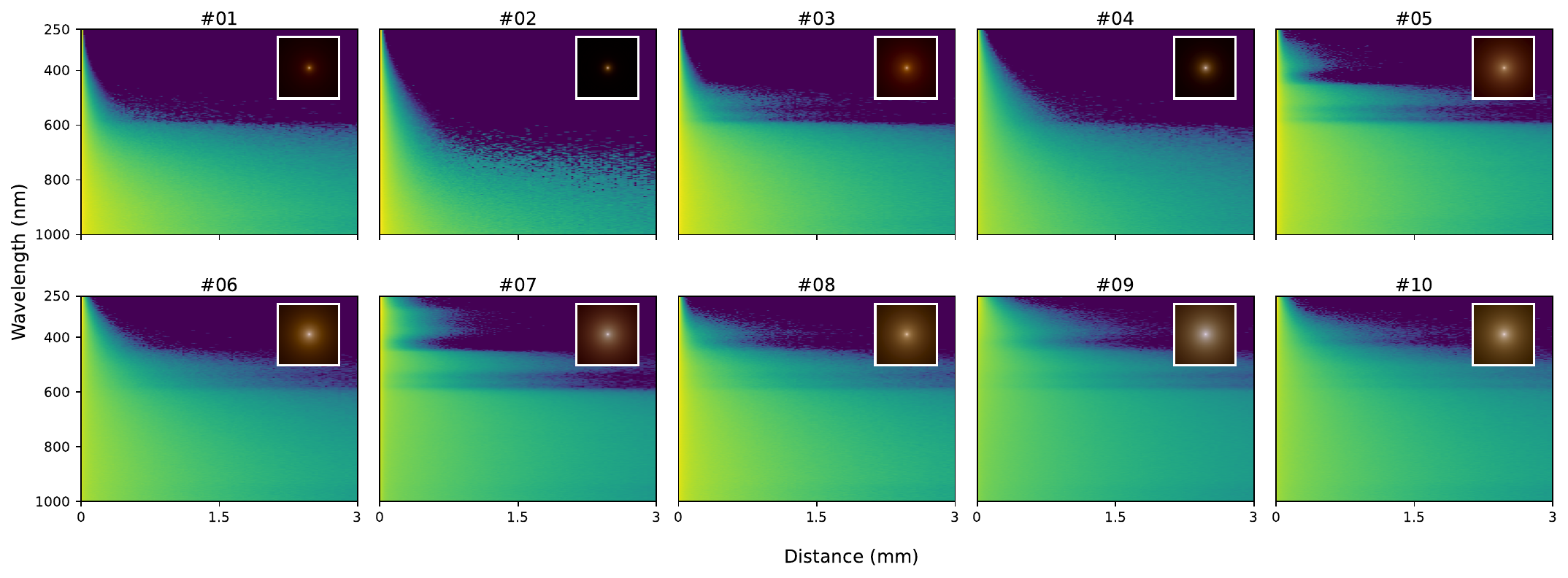}
\end{minipage}

\caption{
\textbf{Left:} Spectral reflectance $R_d(\lambda)$ for 10 representative
skin tones. Melanin dominates UV absorption; haemoglobin creates features
at 415, 542, and 577\,nm. These spectral signatures simultaneously
determine both colour and translucency---the coupling our method
exploits.
\textbf{Right:} Spectral diffusion profiles for representative skin
tones. Main panels: spectral profile heatmap (wavelength $\times$ radial
distance, log scale), showing 3--4 orders of magnitude dynamic range.
Insets: 2D spatial distribution of exiting photons in visible range/RGB
(zoomed to $\pm$2\,mm). Short wavelengths (UV/blue) are confined near the
entry point; long wavelengths (red/NIR) spread broadly.}
\label{fig:diffusion_and_reflectance}
\end{figure}

\paragraph{GPU-accelerated Monte Carlo simulation.}
We developed a GPU Monte Carlo photon-transport simulator for two-layer skin tissue, with wavelength-batched evaluation: All 376 wavelengths (250--1000\,nm, 2\,nm steps) are propagated simultaneously.  For every (skin tone, wavelength) pair we record the total diffuse reflectance $R_d(\lambda)$ and the spatially-resolved radial diffusion profile $R_d(r,\lambda)$ in 512 logarithmically spaced bins out to $r_\text{max}{=}5\,$cm (Fig.~\ref{fig:diffusion_and_reflectance}), with CDF-based conservative resampling. This allows a 4$\times$ and 2$\times$ lower RMSR than uniform binning and kernel density estimation at 10$^5$ photons. 
Unlike prior spectral skin datasets \citep{aliaga2023hyperspectral} that store only hemispherical reflectance, our simulation captures the full spatial distribution of exiting photons, which is necessary for translucency characterisation.

\paragraph{Dataset.}
We generated 25\,000 training and 5\,000 validation skin tones at $10^6$ photons per wavelength on dual A100-80\,GB GPUs. Training parameters are sampled with Halton quasi-random sequences stratified by skin properties, while the validation set is sampled using uniform random sampling. This is a $20{\times}$ reduction from the 600\,000-tone dataset required by \citet{aliaga2023hyperspectral}, enabled by the staged decoder-first training pipeline (Section~\ref{sec:prediction}) that converges with fewer training samples.

\section{Characterizing Subsurface Scattering}\label{sec:characterisation}
The core challenge is to reduce the profiles computed via Monte Carlo to an \emph{effective} set of optical parameters $(\alpha,\,\sigma_t,\,g)$ compatible with path tracers. We will approach this problem by optimizing these parameters to get the best fit possible.

The common practice would be to homogenize the layered medium using a single one. However, as we show in \Cref{fig:k_progression_strip}, this single-medium approach is not expressive enough to capture both the profile shape and the reflectance. Instead, inspired by previous diffusion-based mixture-of-lobes subsurface scattering models \citep{jimenez2015separable,deon2007efficient,christensen2018renderman}, we overcome this problem by using a mixture of $K$ scattering media. The key difference with previous approaches is that this mixture is still expressed in terms of scattering media, and thus each lobe can be computed using the random walk machinery of modern path tracers. 
In the following, we first define our model. Then, we describe our fitting procedure.

\subsection{Subsurface Scattering by Mixtures-of-Media}
\newcommand{\px}{\mathbf{x}}
\newcommand{\py}{\mathbf{y}}
\newcommand{\pf}{\text{pf}}

The radiance scattered out by a translucent surface at point $\px$ in direction $\omega$ can be modeled as
\begin{equation}
    L_o(\px,\omega) = \int_\mathcal{A} \int_{\mathcal{S}^2} L_i(\py \leftarrow \omega_i)\cdot S\left((\px,\omega) \leftarrow (\py,\omega_i)\right) \cdot \cos\theta_i\, d\omega_i d\py, 
\label{eq:rendering}
\end{equation}
where $\mathcal{A}$ and $\mathcal{S}^2$ are the integration area and the unit sphere, $L_i$ is the incident radiance, $\cos\theta_i$ is the Lambert's foreshortening, and $S(\cdot,\cdot)$ is the bidirectional subsurface scattering distribution function (BSSRDF). Note that we omit the spectral dependence, but both radiance and the BSSRDF depend on wavelenght $\lambda$. The BSSRDF is defined as a path integral \citep{veach1998robust} over the space of paths $\bar{\mathbf{x}}\in\Omega$ as
\begin{equation}
S\left((\px,\omega) \leftarrow (\py,\omega_i)\right) = \int_\Omega f_\text{SSS}(\bar{\mathbf{x}}) d\mu(\bar{\mathbf{x}}),
\label{eq:sss}
\end{equation}
with $f_\text{SSS}(\bar{\mathbf{x}})$ the contribution function of path $\bar{\mathbf{x}}=(\py,\px_1'..\px_{k}',\px)$ with length $k+2$, which for an homogeneous medium is defined as (in the inner product we use $\px_0'=\py$ and $\px_{k+1}'=\px$ for simplicity)
\begin{align*}
f_\text{SSS}(\bar{\mathbf{x}}) & =f_r(\omega_i\to\py\to\px_1')\cdot T(\py\to\px_1') \\
& \cdot \prod_{j=1}^k \left(T(\px_j'\to\px_{j+1}') \cdot \pf(\px_{j-1}'\to\px_j'\to\px_{j+1}')\right) \\
& \cdot T(\px_k'\to\px) \cdot f_r(\px_k'\to\px\to\omega)
\end{align*}
with $f_r$ the BTDF at the boundary of the medium, $T(\cdot,\cdot)$ the transmittance between two points, and $\bar{\pf}(\cdot)=\alpha\cdot\sigma_t\cdot\pf(\cdot)$ the phase function multiplied by albedo and extinction coefficient. Again, both transmittance and optical parameters depend on $\lambda$. While diffusion-based approaches approximated $S(\cdot,\cdot)$ using closed-form formulas or mixture-based approximations, most modern path tracers usually solve it by means of Monte Carlo random walks. 
Under normal incidence, and approximating the BTDFs at entry and exit points, the BSSRDF can be described radially as
\begin{equation}
S\left((\px,\omega) \leftarrow (\py,\omega_i)\right) = R_d(|\px-\py|) = R \cdot P_d(|\px-\py|),
\label{eq:radsss}
\end{equation}
with $R_d(|\px-\py|)$ the subsurface profile, $R=\int_\mathcal{A} \int_{\mathcal{S}^2} L_o(\px,\omega) d\omega \px$ under white unit incoming radiance, and $P_d(|\px-\py|)=R_d(|\px-\py|)/R$ the normalized subsurface profile.

\begin{figure}[t]
\centering
\includegraphics[width=\textwidth]{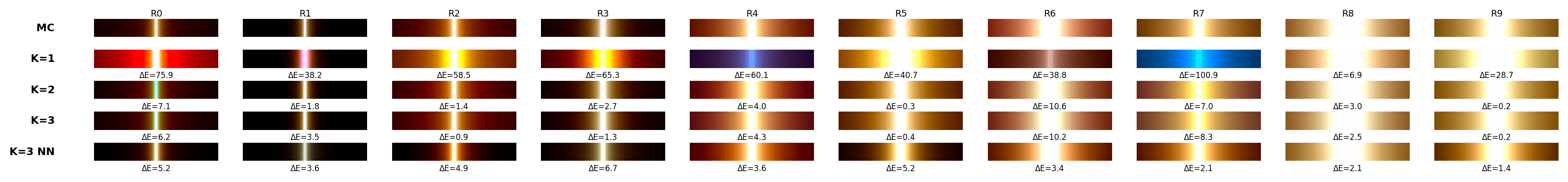}
\caption{Visual K-progression on the 10 representative tones.
Top to bottom: MC ground truth, joint-optimised $K{=}1$, $K{=}2$, $K{=}3$,
and the $K{=}3$ neural decoder (paper model).  Per-cell annotations show
CIE~$\Delta E_{76}$ vs.\ MC ground truth.
$K{=}1$ cannot decouple shape from amplitude, producing monochromatic
streak artefacts.  $K{=}2$ collapses the error from $\Delta E\!=\!34.7$ to
$3.81$; $K{=}3$ flatlines at $3.79$.  The NN row (bottom) surpasses the
joint-opt ceiling ($3.19$ vs.\ $3.79$; Tab.~\ref{tab:mc_gate}).
}
\label{fig:k_progression_strip}
\end{figure}

As we show in \Cref{fig:k_progression_strip}, a single homogeneized medium fitted to the reference profile only roughly approximates the two-layers profile. Instead, we introduce additional complexity in \Cref{eq:sss} using a $K$-lobe mixture of media, where each medium $k$ is defined by its own homogeneized optical parameters $(\alpha_k, \sigma_{t,k}, g_k)$ plus a weight $w_k$, resulting in
\begin{equation}
S\left((\px,\omega) \leftarrow (\py,\omega_i)\right) = \sum_{k=1}^K w_k\int_\Omega f_\text{SSS,k}(\bar{\mathbf{x}}) d\mu(\bar{\mathbf{x}}),
\label{eq:sss}
\end{equation}
where $f_\text{SSS,k}(\bar{\mathbf{x}})$ is the throughput of the path $\bar{\mathbf{x}}$ in medium $k$, and $\sum_{k=1}^K w_k=1$. This can be trivially implemented on a random-walk-based subsurface integrator, by simply randomly selecting a medium before start tracing the random walk. Note that this is different from blending the optical parameters of $K$ media together.

\subsection{Fitting Subsurface Scattering}
Given the large scale of our subsurface profiles dataset, optimizing with naive differentiable rendering is untracktable. Instead, we leverage a more efficient approach, that allows us to build our large scale mapping between our biophysical manifold of skin tones and their best-fit $K$-media mixture. At the core of our optimization, we use a precomputed lookup table (LUT) that depends on the albedo and anisotropy only, and that allows us to avoid Monte Carlo simulations during fitting.

\paragraph{Scale-independent lookup table}\label{sec:lut}
For a homogeneous medium, \citet{chiang2016practical} showed that the normalized subsurface profile shape $R_d(r;\,\alpha,\,\sigma_t)$ depends only on $\alpha$ (and~$g$); extinction $\sigma_t$ only rescales the radial axis:
\begin{equation}\label{eq:separability}
  P_d(r\,|\,\alpha,\sigma_t) \;=\;
  \sigma_t^{2}\;\cdot\;P_d\!\left(\sigma_t \cdot r\,|\,\alpha,\sigma_t^\text{ref}{=}1\right).
\end{equation}
This separability does not hold for multi-layer media such as skin, where depth-dependent absorption couples shape to extinction. Nevertheless, the homogeneous LUT provides a useful building block: A LUT $\mathcal{P}_d(r\,|\,\alpha,\,g)$ defined over the single scattering albedo and anisotropy is independent of both $\sigma_t$ and wavelength, and thus can be densely populated for accelerating the optimization procedure. We precompute an $800{\times}800$ LUT using $10^6$ photons per $(\alpha, g)$-combination. Since for a flat semi-infinite media is scale invariant, We also precompute a 2D LUT $\mathcal{R}(\alpha,g)$ for the reflectance $R$ with the same dimensions as $\mathcal{P}_d$.

With our pair of LUTs with precomputed subsurface profiles and reflectances, we can quickly compute the subsurface profile of $K$-media mixture as
\begin{equation}\label{eq:mixture}
  R_d^\text{fit}(r) = \sum_{k=1}^{K} w_k\cdot \mathcal{R}(\alpha_k,g_k) \cdot \mathcal{P}(r\,|\,\alpha_k,g_k,\sigma_{t,k}),
\end{equation}
with the LUTs accessed using bilinear and trilinear interpolation, respectively. Note that linear interpolation is differentiable, making this approach suitable for optimization. 

\paragraph{Joint shape-and-amplitude loss function}
We jointly fit all parameters for both profile shape and spectral
reflectance. For each skin tone and wavelength we minimize the loss
\begin{equation}\label{eq:joint_loss}
  \mathcal{L}_\text{joint} = \mathcal{L}_\text{shape} + \lambda_R\cdot\mathcal{L}_R,
\end{equation}
where $\mathcal{L}_\text{shape}$ and $\mathcal{L}_R$ measure the error loss of the of the normalized subsurface profile and the reflectance, against their Monte Carlo references $P_d^\text{MC}(r)$ and $R_\text{MC}$ respectively. The two losses are defined as
\begin{align}
  \mathcal{L}_\text{shape} &= \text{RMSE}\bigl(\log\hat{P_d}(r),\;\log P_d^\text{MC}(r)\bigr),
  \label{eq:shape_loss}\\
  \mathcal{L}_R &= \bigl(\hat{R} - R_\text{MC}\bigr)^2,
  \label{eq:refl_loss}
\end{align}
with $\hat{P_d}(r) = \sum_k w_k\cdot \mathcal{P}(r\,|\,\alpha_k,g_k,\sigma_{t,k})$, $\hat{R} = \sum_k w_k\cdot\mathcal{R}(\alpha_k, g_k)$ and $\lambda_R = 5$. 
The log-space shape comparison for the scattering profile equalizes the importance of the peak and tail.

\paragraph{Optimization.}\label{sec:levels}
We optimize using Adam (lr=$10^{-2}$, 10\,000 steps with cosine schedule) per tone per wavelength. We initialize training by using $K=1$ fitting only the profile via the LUT.
For $K{=}2$, we jointly
optimize all seven parameters (only optimize a single weight value $w$, since the other can be computed from it) 
over 3000 steps, achieving mean shape and reflectance errors of 0.51 and 5.0\% respectively for 25K tones.
For $K{=}3$, we expand the dimensionality by splitting the dominant component and optimise all 11 parameters per wavelength over 10\,000 steps, reaching mean errors of 0.196 and 3.7\% for profile shape and reflectance, respectively. 
Importantly, the $\sigma_t$ parameterisation is \emph{unbounded} ($\sigma_t = \exp(\text{raw})$, no clamping), allowing the optimiser to discover the full physical range.
All 25\,000 training tones $\times$ 376 wavelengths are batched on GPU (46\,h on a single A100 with 80\,GB).


\section{Joint Albedo-Translucency Prediction}\label{sec:prediction}

The joint-optimized targets from \Cref{sec:characterisation} provide
per-tone, per-wavelength high-quality fits, but their computation requires 46\,h of optimization for 25\,000 tones. For a more practical continuous approach, we distil this into a single-forward-pass neural decoder, that takes as input an RGB skin tone $A_\text{RGB}$, and produces spectral subsurface parameters $\hat{\boldsymbol{\theta}}$ for our $K$-media mixture (we set $K=3$), directly usable in production renderers with random-walk and diffusion-based subsurface scattering.

\subsection{Architecture: Chained Decoder}\label{sec:chained_decoder}

The key architectural decision is to condition the translucency decoder on the full predicted spectral reflectance, not solely on the 5D biophysical parameters. This provides a 376-dimensional spectral context that disambiguates the otherwise ill-posed inverse mapping. We make the $\lambda$-dependence explicit again, to emphasize the spectral nature of our decoder. 

Our neural architecture consists of three chained components: a) An encoder $\text{Enc}(A_\text{RGB})\to\hat{\mathbf{p}}$ that takes the input RGB skin tone albedo $A_\text{RGB}\in\mathbb{R}^3$ and outputs a set of estimated biophysical parameters $\hat{\mathbf{p}}\in\mathbb{R}^5$; b) a reflectance decoder $\text{Dec}_R(\hat{\mathbf{p}})\to \hat{R}_d(\lambda)$ that maps $\hat{\mathbf{p}}$ to an spectral reflectance $\hat{R}_d(\lambda)\in\mathbb{R}^{376}$; and c) a final translucency decoder $\text{Dec}_T(\mathbf{p},\hat{R}_d(\lambda))\to\hat{\boldsymbol{\theta}}(\lambda)$ that takes both $\hat{\mathbf{p}}$ and $\hat{R}_d(\lambda)$ and generates the final spectral optical parameters for the $K$-media mixture 
\begin{equation}\label{eq:decoder_output}
  \hat{\boldsymbol{\theta}}(\lambda) = \bigl\{\hat\alpha_k(\lambda),\;\hat{g}_k(\lambda),\;\hat\sigma_{t,k}(\lambda),\;\hat{w}_k(\lambda)\bigr\}_{k=1}^{K} \in \mathbb{R}^{376\times 4K}.
\end{equation}

The architecture of the encoder $\text{Enc}(\cdot)$ and reflectance decoder $\text{Dec}_R(\cdot)$ is the same as the one proposed by \citet{aliaga2023hyperspectral}. 
The translucent decoder $\text{Dec}_T(\cdot)$ is a 4-layer MLP, with input dimension 381, hidden dimensions 768 with tanh activation functions, and output $376\times 12$ with sigmoid activation. We enfore that the output weights $\sum_k \hat{w}_k = 1$ via softmax
over the raw logits.



\subsection{Frozen Staged Training}\label{sec:staged_pipeline}

The three components ($\text{Enc}(\cdot)$, $\text{Dec}_R(\cdot)$, and $\text{Dec}_T$) are trained sequentially, each freezing its predecessors. This staged design prevents gradient interference between components and ensures that each module converges to a stable representation before the next is built upon it.

\textbf{Stage~1 (Decoder $\text{Dec}_R(\cdot)$)} trains the mapping
$\mathbf{p} \to \hat{R}_d(\lambda)$ with a spectral L1 loss
(300 epochs, spectral L1 $= 0.001$). Once frozen, this decoder serves as a differentiable neural spectral renderer, which replaces costly Monte Carlo forward evaluation in subsequent stages and enables a $20{\times}$ trainihg dataset reduction, as noted in Section~\ref{sec:model}.

\textbf{Stage~2 (Encoder $\text{Enc}(\cdot)$)} trains the inverse mapping from RGB albedo $\mathbf{A}_\text{RGB} \to \hat{\mathbf{p}}$, using the frozen $\text{Dec}_R(\cdot)$ to provide spectral reconstruction from the estimated parameters. We optimize a color reconstruction loss: $\mathcal{L}_{\text{enc}} = | \mathbf{A}_{\text{RGB}} - \gamma \cdot \text{CMF}(\text{Dec}_R(\text{Enc}(\mathbf{A}_{\text{RGB}}))) |_1$, without a parameter supervision. 
The encoder additionally predicts a scalar exposure $\gamma$ that relates the input albedo to the unit-exposure reflectance prediction:
\begin{equation}\label{eq:exposure}
  \gamma = \frac{Y(\mathbf{A}_\text{RGB})}{Y(\hat{\mathbf{A}})},
\end{equation}
where $Y(\cdot)$ denotes CIE relative luminance
and $\hat{\mathbf{A}}$ is the color reconstructed from the predicted spectrum via D65 color matching. This analytical definition prevents the exposure channel from absorbing residuals that other parameters fail to explain, a known failure mode of unconstrained latent representations \citep{aliaga2023hyperspectral}.

\textbf{Stage~3 (Decoder $\text{Dec}_R(\cdot)$)} trains the translucency prediction component with frozen encoders and decoders, We detail the training losses and procedure in \Cref{sec:decr}.

Empirically, unfreezing any preceding component during a later stage causes up to 40\% regression in error $\Delta E$. Note that the coupling between stages is strictly feed-forward.

\subsection{Training the translucency prediction decoder}
\label{sec:decr}
We train the translucency prediction decoder by minimizing the following loss:
\begin{equation}\label{eq:total_loss}
  \mathcal{L} = \mathcal{L}_\text{param}
  + \lambda_R\,\mathcal{L}_\text{refl}
  + \lambda_\text{smooth}\,\mathcal{L}_\text{smooth},
\end{equation}
with three terms:
\begin{align}
  \mathcal{L}_\text{param} &= \frac{1}{|\mathcal{C}|}
    \sum_{c} \phi_c \cdot \text{Huber}\bigl(\hat{\theta}_c,\;\theta_c^\star\bigr),
    \label{eq:lparam}\\
  \mathcal{L}_\text{refl} &= \bigl\|
    \hat{R}_\text{LUT}(\lambda) - R_\text{MC}(\lambda)
    \bigr\|_1,
    \label{eq:lrefl_lut}\\
  \mathcal{L}_\text{smooth} &= \bigl\|
    \nabla_\lambda\hat{\boldsymbol{\theta}}\bigr\|_1,
    \label{eq:lsmooth}
\end{align}
where $\theta_c^\star$ are the joint-optimised targets (\Cref{eq:joint_loss}), and $\nabla_\lambda\hat{\boldsymbol{\theta}}$ is the finite-difference spectral gradient of the predicted parameters.  The Huber loss in $\mathcal{L}_\text{param}$ provides L1-like robustness to outlier tones
while remaining differentiable at zero.

We set $\lambda_R{=}15$, which we found empirically optimal (lower values under-weight reflectance causing colour drift, while higher values produce gradient instability at the slab-correction boundary) and $\lambda_\text{smooth}{=}0.50$.
The $\alpha$ channels receive doubled weight ($\phi_\alpha = 2$) in $\mathcal{L}_\text{param}$ to reflect their dominant role in determining reflectance.

\paragraph{Reflectance loss} We introduce a color loss term $\mathcal{L}_\text{refl}$ since parameter-space supervision alone produces a $\Delta E = 11$ color error. This is because the reflectance $R_d$ is a highly nonlinear function of $(\alpha, g)$, and small parameter errors compound into visible colour shifts.  
We add a differentiable LUT loss that directly penalises spectral reflectance error through the precomputed LUT, computing $\hat{R}_\text{LUT}(\lambda) = \sum_{k=1}^{K} \hat{w}_k(\lambda)\cdot \mathcal{R}\bigl(\hat\alpha_k(\lambda),\, \hat{g}_k(\lambda)\bigr)$. This reduces colour error from $\Delta E = 11$ (parameter regression alone) to imperceptible levels ($\Delta E = 0.062$).


\paragraph{Extinction-aware reflectance coupling}
Due to similarity relations (\Cref{eq:separability}), the reflectance is independent of $\sigma_t$, and thus the color loss has zero gradients w.r.t extinction for a fixed pair $(\alpha,g)$. 
%
We avoid this by adding a bounded $\sigma_t$-dependent correction via a learned
slab-depth factor, which modifies the predicted reflectance as 
\begin{equation}\label{eq:slab_factor}
  \hat{R}_k(\lambda) = \mathcal{R}\bigl(\hat\alpha_k, \hat{g}_k\bigr)
  \cdot \bigl(1 - e^{-C\cdot \hat\sigma_{t,k}(\lambda)\cdot c(\lambda)}\bigr),
\end{equation}
where $c(\lambda) \in \mathbb{R}^{376}$ is a learned per-wavelength slab
depth (initialised at 0.30, jointly optimized with the decoder weights),
and $C{=}3$ is a fixed constant. The factor $(1 - e^{-C\,\sigma_t\,d})$ saturates to unity at high $\sigma_t$, recovering the semi-infinite LUT regime, while providing a bounded non-exploding gradient at moderate $\sigma_t$ values where extinction affects the effective slab optical thickness.
Without this correction, the decoder has no gradient signal on $\sigma_t$ through the reflectance loss and the error plateaus at $\Delta E \approx 7.6$.  With the learned $c(\lambda)$, the error improves to $\Delta E = 3.19$ (Section~\ref{sec:mc_gate}).

\paragraph{Inductive biases.}
Two properties of the neural decoder act as implicit regularizers that enable it to surpass the per-tone joint optimization.  First, spectral smoothness $\mathcal{L}_\text{smooth}$ prevents wavelength-to-wavelength oscillations at degenerate basins where the subsurface profile/reflectance tradeoff is ambiguous. Second, shared weights across all 25\,000 training tones enforce cross-tone consistency---the network cannot overfit to noise in individual tones.  Together, these produce spectrally coherent parameters that integrate to better rendered color than independent per-tone optimization.

\section{Validation}\label{sec:validation} 

We validate our method along three axes: (1)~Rendering quality, (2)~spectral colour accuracy, and (3)~subsurface profile shape fidelity across $K$-components.

\paragraph{Reflectance quality}\label{sec:mc_gate}
Our primary metric is end-to-end rendering quality: Predicted $K$-component
parameters are fed into a full Monte Carlo random-walk renderer on a
semi-infinite slab (500\,000 photons per wavelength), and the resulting
spectral image is compared against the joint-optimized reference rendered
through the same pipeline, as described in \Cref{sec:model}, with the key difference that each photon randomly selects a medium from the $K$-media mixture. 
Predicted and reference parameters are rendered independently; the resulting D65/sRGB images are compared via CIE~$\Delta E_{76}$ (Lab).  This is 
intentionally stricter than profile-distance metrics: Parameter inaccuracies
that average out in profile space are amplified through the per-component
scattering integral.

\begin{table}[t]
\centering
\caption{Rendering error ($\Delta E_{76}$ on the visible band) on 10
  representative skin tones.  The $K{=}3$ NN decoder surpasses the
  joint-optimized ceiling ($\Delta E\!=\!3.19$ vs.\ 3.79), which demonstrates the that the whole pipeline provides a better landscape for optimization.}
\label{tab:mc_gate}
\small
\begin{tabular}{@{}l c c l@{}}
\toprule
Configuration & Rep $\Delta E$ & vs K=1 & Notes \\
\midrule
$K{=}1$ (joint-opt) & 51.40 & --- & Single-medium limit \\
$K{=}2$ (joint-opt) & 3.80 & $13.5{\times}$ & Fitting ceiling \\
$K{=}3$ (joint-opt) & 3.79 & $13.6{\times}$ & $\approx K{=}2$ (structural only) \\
\midrule
$K{=}3$ NN (paper model) & 3.19 & $16.1{\times}$ & seed=123, smooth=0.50, $d(\lambda)$ \\
\quad vs.\ joint-opt & $-0.60$ & & Surpasses ceiling \\
\bottomrule
\end{tabular}
\end{table}

Table~\ref{tab:mc_gate} reveals that $K=1$ cannot simultaneously match both the subsurface profile and the reflectance, while $K=2$ solves this completely. 
$K{=}3$ provides marginal fit-quality improvement $K{=}3$, with only a structural structural contribution (Section~\ref{sec:k3_interpretation}).
The neural decoder achieves $\Delta E = 3.19$ on the 10 representative tones, surpassing the joint-optimised ceiling of 3.79.  On the combined 20-tone gate (10 representative + 10 outliers) 
it scores $\Delta E = 3.51$.  The negative distillation gap ($-0.60$) confirms that a single network with shared weights can replicate---and improve upon---per-tone independent optimisation.


\begin{table}[t]
\centering
\caption{Banded fidelity metrics (joint-opt, 10 representative tones).
  $K{=}3$ provides no improvement over $K{=}2$ in any band.}
\label{tab:banded}
\small
\begin{tabular}{@{}l ccc cc@{}}
\toprule
Band & Shape $K{=}1$ & $K{=}2$ & $K{=}3$ & $K{=}1{\to}2$ & $K{=}2{\to}3$ \\
\midrule
UV (250--400\,nm)  & 14.97 & 14.80 & 14.81 & 1.1\%  & 0.0\% \\
Vis (400--700\,nm) & 2.64  & 2.20  & 2.16  & 16.9\% & 1.8\% \\
NIR (700--1000\,nm)& 2.30  & 2.28  & 2.27  & 0.6\%  & 0.4\% \\
\midrule
SAM Vis (rad) & 0.69 & 0.14 & 0.13 & 79.9\% & 2.9\% \\
\bottomrule
\end{tabular}
\end{table}
\paragraph{Spectral band fidelity}
To verify that $K{=}3$'s structural contribution is not masked by
band-specific effects, we compute shape error (log-space RMSE of predicted
vs.\ MC profiles) and spectral angle (SAM) per spectral band in \Cref{tab:banded}. 
The $K{=}1{\to}K{=}2$ improvement concentrates in the visible band (79.9\%
SAM reduction), corresponding to haemoglobin absorption features in the
540--577\,nm window that a single medium cannot decouple from melanin.
UV and NIR errors are dominated by MC photon noise at low reflectance
($R(\lambda) \sim 0.01$--$0.05$) and are $K$-independent.

\paragraph{Subsurface profile fidelity}
Finally, we analyze the error introduced on the subsurface profile reconstruction. On the full 5\,000-tone validation set (376 wavelengths each), $K{=}2$ reduces mean profile shape error by 67.3\% relative to $K{=}1$ (0.059\,$\to$\,0.019, log-space RMSE). 

\begin{figure}[t]
\centering
\includegraphics[width=0.95\columnwidth]{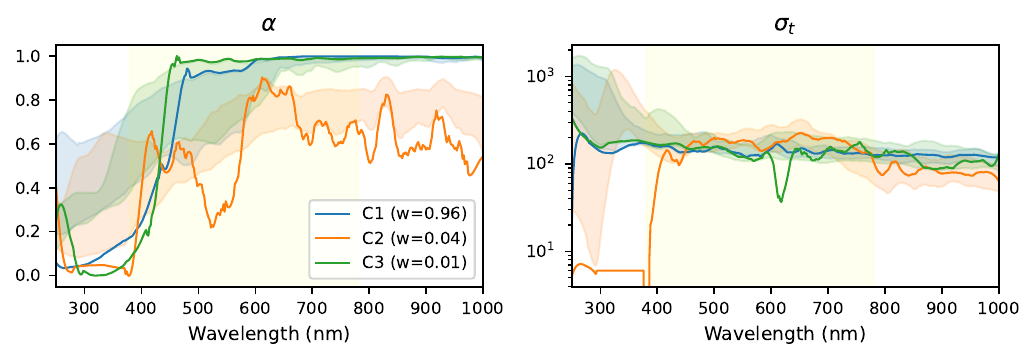}
\caption{Predicted spectral scattering parameters ($K{=}3$ decoder),
pooled across all subjects.  Top: single-scattering albedo
$\alpha(\lambda)$ per component (mode $\pm$ IQR).  Bottom: extinction
$\sigma_t(\lambda)$.  Haemoglobin absorption features (415, 542,
577\,nm) and melanin UV absorption are captured as per-wavelength
structure that scalar parameters cannot represent.  Component weights
shown in inset.}
\label{fig:spectral_params}
\end{figure}
\paragraph{Spectral SSS parameters}
Figure~\ref{fig:spectral_params} shows the predicted per-wavelength parameters pooled across all subjects.  All expected physical trends are recovered: $\alpha$ decreases with melanin content (darker skin absorbs more), $\sigma_t$ increases (shorter mean free path), and haemoglobin absorption bands appear as spectral features in both $\alpha$ and $\sigma_t$.


\section{Results}\label{sec:results}
We implement our model in Mitsuba 3 \citep{jakob2022dr}, as a new random-walk-based subsurface integrator. The only (minimal) modification is at the begining of the random walk, where we randomly select the medium to traverse from the $K$-media mixture, and adjust the pdf accordingly. We sample each medium based on its weight $w_k$, though other approaches could be explored. 


\begin{figure*}[t]
\centering
\includegraphics[width=\textwidth]{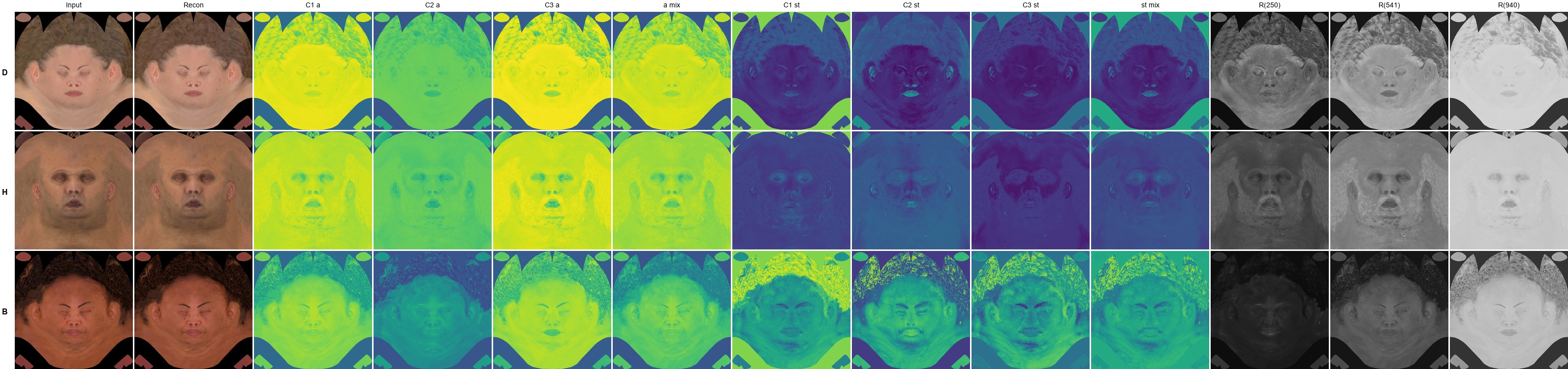}
\caption{SSS parameter reconstruction and spectral reflectance for three
subjects spanning light to dark skin (top to bottom: D, H, B).
Per-component single-scattering albedo $\alpha\in[0,1]$ and extinction
$\sigma_t$ (viridis, log$_{10}$ scale, globally normalised;
C1: $[80,\,901]$, C2: $[73,\,3555]$, C3: $[86,\,5147]$\,cm$^{-1}$)
predicted by the $K{=}3$ decoder.  Right three columns show spectral
reflectance at 250\,nm (UV), 541\,nm (haemoglobin), and 940\,nm (NIR),
confirming wavelength-dependent translucency.  Input RGB albedo and
reconstructed colour shown at left.}
\label{fig:sss_reconstruction}
\end{figure*}

\Cref{fig:teaser} compares renders using RGB albedo inversion \citep{wrenninge2017path} with our automatic prediction. Wrenninge's fixed scattering distance produces uniform translucency regardless of skin tone or needs to be manually adjusted (like in the examples shown in the figure); our method correctly modulates translucency with chromophore content, producing darker skin with shorter mean free paths and lighter skin with broader subsurface spread. In addition, our method provides fully spectral optical parameters beyond the visible spectrum, while previous albedo inversion methods would require manual tuning of extinction and anisotropy. 

Figure~\ref{fig:sss_reconstruction} shows the full per-component
parameter decomposition for three subjects spanning light to dark skins, including spectral reflectance at representative wavelengths. The reconstructed color (second column) confirms spectral accuracy of the whole pipeline.

\section{Discussion}\label{sec:discussion}
\paragraph{The anisotropy degree of freedom is essential.}
Optimizing based on the isotropic assumption of \citet{chiang2016practical} has a significant error ($\Delta E=0.5$), compared to include this anisotropy ($\Delta E=0.019$). This confirms the need of using anisotropic phase functions (and albedo inversion) when rendering skin. 

\begin{figure}[t]
\centering
\includegraphics[width=0.95\columnwidth]{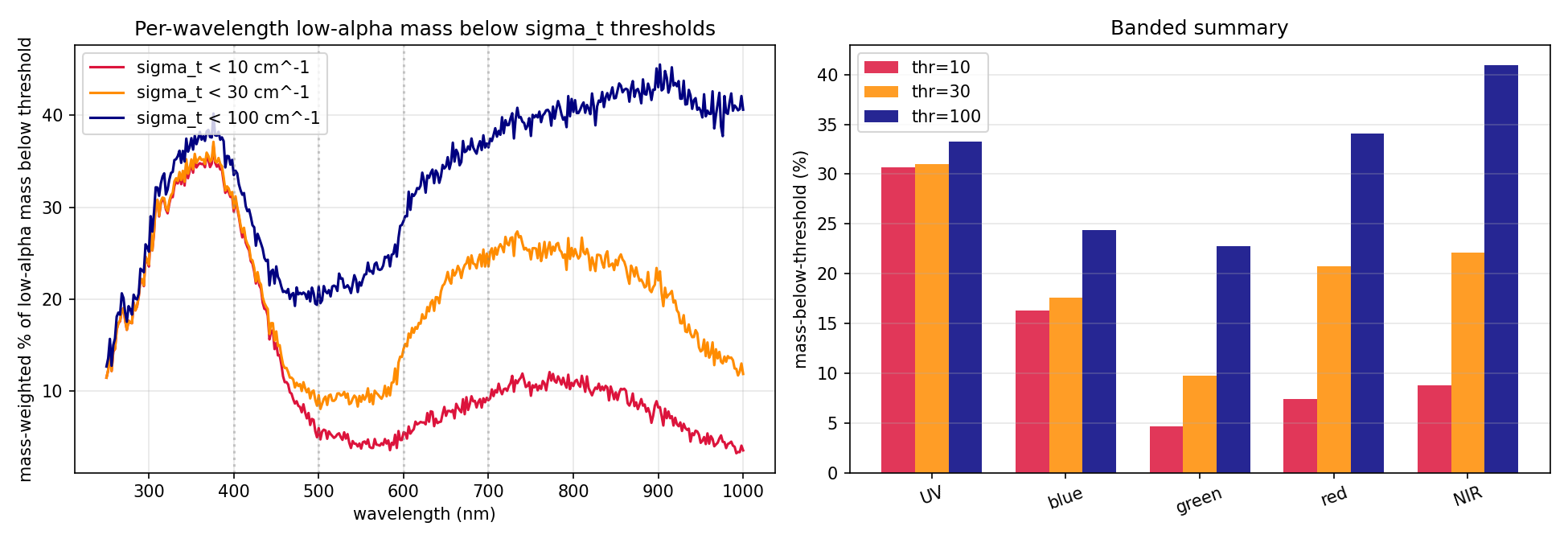}
\caption{Per-wavelength $\sigma_t$ mass distribution for the
low-$\alpha$ component (C3).  UV and blue bands concentrate 30.7\% and
16.3\% of mass in the boundary-equilibration regime
($\sigma_t < 10$\,cm$^{-1}$), vs 8.8\% in NIR---a $2.7{\times}$
concentration ratio consistent with epidermal melanin absorption
dominating short-wavelength paths.}
\label{fig:sigma_t_distribution}
\end{figure}
\begin{figure}[t]
\centering
\includegraphics[width=0.95\columnwidth]{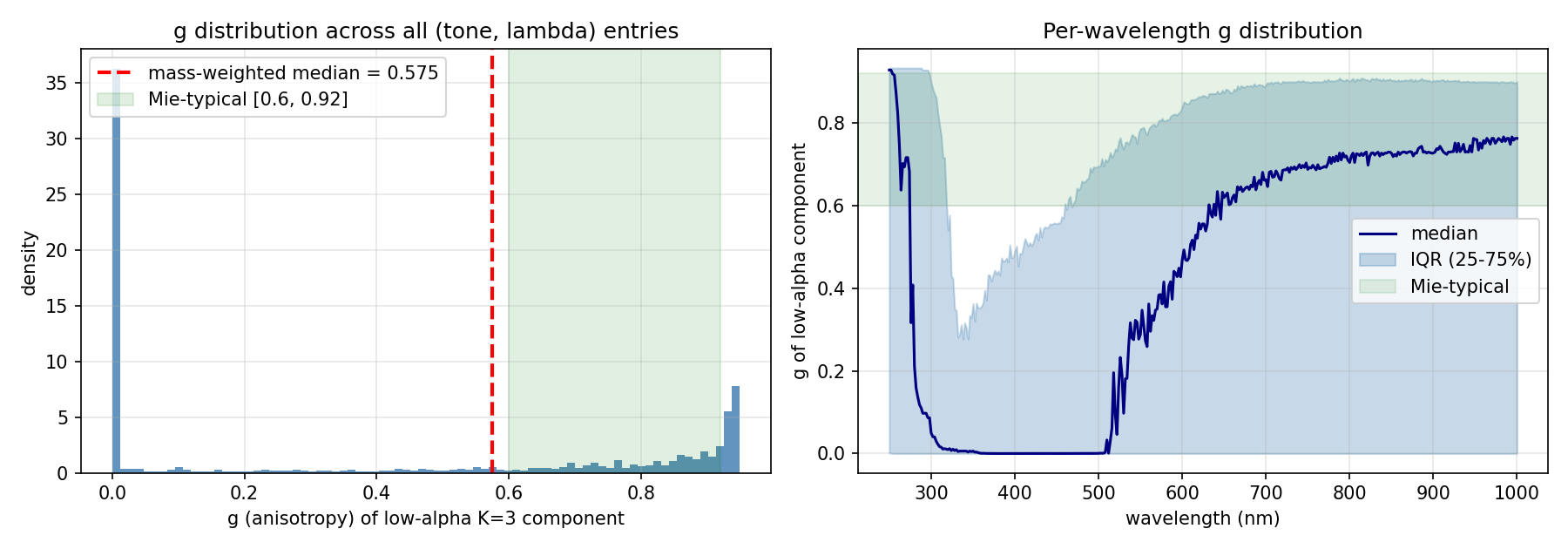}
\caption{Phase-function anisotropy $g$ bimodality for the
low-$\alpha$ component (C3).  The bimodal distribution (modes at
$g{\approx}0$ and $g{\approx}0.87$) independently confirms the
few-scatter interpretation: when photons undergo few events, the phase
function is poorly constrained by reflectance.}
\label{fig:g_bimodality}
\end{figure}

\paragraph{$K{=}3$ physical decomposition.}\label{sec:k3_interpretation}
For $K=3$ we observe that the two high-$\alpha$ components (median $\alpha \approx 0.96$) remain in the scattering-dominated regime, which is consistent with bulk dermal transport.  The third component (median $\alpha \approx 0.34$, weight~0.35) stays at the diffusion boundary: 23.1\% of its mixture-weighted energy extends into the boundary-equilibration regime ($\sigma_t < 10$\,cm$^{-1}$), and concentrating its weigth in the concentrating in UV and blue bands ($2.7{\times}$ over NIR; Fig.~\ref{fig:sigma_t_distribution}): This is consistent with epidermal melanin absorption terminating short-wavelengths paths before diffusive equilibrium. 
This is also confirmed by the bimodal $g$ distribution of this third component (Fig.~\ref{fig:g_bimodality}): Too few events for angular redistribution to constrain the phase function.

\paragraph{Limitations and future work.}
Our fitting procedure assumes a smooth Fresnel boundary illuminated by a cosine-weighted lobe, collapsing the full angular dependence of incident light into a single normal-incidence profile. This approximation has two consequences: First, surface roughness and micro wrinkles are not accounted for. Second, in the two-layer regime with high melanin, the epidermis acts as a thin, highly absorbing slab through which photons must pass twice; the fitted effective-medium components compensate with both large $g$ and large $\sigma_t$, but this causes that low-order scattering events dominate the near-field part of the subsurface profile, but are understimated by the cosine-weighted entry assumption. The main consquence is that, for grazing angles where the radial assumption does not hold, the homogeneized medium might be too conservative, producing overly sharp translucency in these regions. A potential alternative for fixing this issue is to account for the incoming light direction as an additional parameter in our inversion, though this would require a significantly larger training set, and a more complex integration within the renderer. 

An additional problem is that, while our technique generalizes well to skin tone, it struggles in areas such as lips which fall out of the manifold of skin tones used for training. In this areas, our predicted parameters result in a highly absorbing medium (see \Cref{fig:teaser}). Incorporating these additional areas into the training set, or training a region-specific inversion would allow to use our methodology in these regions. 

The $K{=}3$ analysis reveals that a non-negligible fraction of mixture-weighted mass operates in the boundary-equilibrium diffusion regime where the scale-independent 2D~LUT is no longer exact (\Cref{sec:k3_interpretation}). Extending the LUT to include explicit $\sigma_t$ dependence would remove this extinction-blindness. 

Finally, our validation relies on Monte Carlo simulation. Face renders on real captured albedos (Fig.~\ref{fig:teaser}) demonstrate generalisation beyond the training distribution. Fitting to measured diffusion profiles from integrating-sphere measurements would provide an independent physical validation axis. Still, evaluating our predicted subsurface profiles on captured in-vivo samples, to fully validate our model, would be an interesting next step.


\section{Conclusion}\label{sec:conclusion}

We have presented a framework for predicting per-wavelength spectral skin translucency from RGB reflectance. This generalizes existing albedo inversion methods by not only by giving the whole set of parameters required for modern random-walk subsurface light transport simulation, but also by providing them in a fully spectral mode, including near-infrared and ultra-violet ranges. 

Our method is based on two main contributions: A mixture-of-media formulation for modeling complex multilayerd scattering media; and a neural inversion method that is capable of extracting optical parameters directly from skin reflectance textures. We demonstrated our methods by integrating it on a spectral path tracer. To our knowledge, this is the first system providing biophysically-grounded per-wavelength subsurface scattering parameters for skin, enabling spectral rendering of translucency from a single RGB albedo.


\section{Acknowledgements}
We want to thank Christophe Hery for initial discussions on the topic. 

\bibliographystyle{assets/plainnat}
\bibliography{paper}

\clearpage
\newpage

\end{document}